\documentclass[
superscriptaddress,
twocolumn,
showpacs,
showkeys,
pra,
floatfix,
]{revtex4-1}

\usepackage{amsmath}
\usepackage{amssymb}

\usepackage{graphicx}
\usepackage{bm,bbm}
\usepackage{color}

\newcommand{\rmw}{\mathrm{w}}
\newcommand{\rmd}{\mathrm{d}}
\newcommand{\rme}{\mathrm{e}}
\newcommand{\rmi}{\mathrm{i}}

\newcommand{\rmK}{\mathrm{K}}
\newcommand{\One}{\openone}

\newcommand{\mt}{\mathrm }
\newcommand{\tauk}{\tau_{\mathrm{k}}}

\newcommand{\omo}{\omega_{\mathrm{o}}}

\newcommand{\h}{\hat}

\newcommand{\la}{\langle}
\newcommand{\ra}{\rangle}

\newcommand{\eg}{\textit{e.g.}}

\newcommand{\Hosc}{H_{\mathrm{o}}}
\newcommand{\Hkosc}{H_{\mathrm{ko}}}
\newcommand{\bM}{\bm{M}}
\newcommand{\bA}{\bm{A}} 
\newcommand{\Eqst}{\overline{E}_{\mt{qst}}}

\begin{document}
\title{Quantum kicked harmonic oscillator in contact with a heat bath}

\author{M. A. Prado}
\affiliation{Departamento de F\'\i sica, Universidad de Guadalajara,
  Blvd. Marcelino Garc\'\i a Barragan y Calzada Ol\'\i mpica, 
  C.P. 44840, Guadalajara, Jalisco, M\' exico.}  
  
\author{P. C. L\' opez V\' azquez}
\affiliation{Departamento de Ciencias B\'asicas, Tecnol\'ogico de Monterrey, Campus Guadalajara,  
Av. General Ram\'on Corona No. 2514, C.P. 45201, Zapopan, Jalisco, M\'exico}
\affiliation{Max-Planck-Institut f\" ur Physik komplexer Systeme, D-01187
  Dresden, Germany}

\author{T. Gorin}
\affiliation{Departamento de F\'\i sica, Universidad de Guadalajara,
  Blvd. Marcelino Garc\'\i a Barragan y Calzada Ol\'\i mpica, 
  C.P. 44840, Guadalajara, Jalisco, M\' exico.}
  
\begin{abstract}
We consider the quantum harmonic oscillator in contact with a finite 
temperature bath, modeled by the Caldeira-Leggett master equation. Applying 
periodic kicks to the oscillator, we study the system in different dynamical 
regimes between classical integrability and chaos on the one hand, and
ballistic or diffusive energy absorption on the other. We then investigate the 
influence of the heat bath on the oscillator in each case. Phase space 
techniques allow us to simulate the evolution of the system efficiently. In 
this way, we calculate high resolution Wigner functions at long times, where 
the system approaches a quasi-stationary cyclic evolution. Thereby, we are able 
to perform an accurate study of the thermodynamic properties of a 
non-integrable, quantum chaotic system in contact with a heat bath. 
\end{abstract}

\maketitle

\section{\label{I} Introduction}

Recently, the ``emergence of thermodynamic behavior within composite quantum 
systems''~\cite{GeMiMa09} has become a very active research field. In this
contribution, we present a numerical study of the implications of ``quantum 
chaos''~\cite{Stoe99,Haake2001} for the emergence of thermodynamic behavior in 
a quantum system coupled to a finite temperature heat bath. 

In open quantum systems, quantum chaos has been investigated first of all on 
the side of the environment. The question was then whether quantum chaos would 
imply special noticeable effects on the central system. This was studied, for
instance, by comparing the effects of a particular quantum chaotic environment
to the ubiquitous collection of harmonic oscillators~\cite{Sre94,Dor97}. 
Similarly, in the spirit of the quantum chaos 
conjecture~\cite{BerTab77,CVG80,Boh95,BluSmi90} Lutz and Weidenm\" uller 
studied an environment modeled by random matrix theory~\cite{LutWei99}. Later on 
random matrix environments have been used to describe decoherence processes 
with applications mainly in decoherence~\cite{EspGas04} and quantum information 
processes~\cite{GS02b,GS03,PGS07,GPKS08,Dav11,CGS14}. However, the first paper
on quantum open systems, derived from a random matrix approach for the 
environment is Ref.~\cite{GeRiFr72} which deals with the description of highly
excited vibrational states in molecules.

However, such studies do not explain how a quantum chaotic environment can 
appear in an experimental setup. Of course one may simply assume that system
and quantum chaotic environment are perfectly isolated from anything else,
but this is a rather unrealistic assumption. In practice, it will be inevitable
that the quantum chaotic environment is in contact with other degrees of 
freedom, not being considered so far, the ``far environment''. In this work, we
assume the far environment to act as a finite temperature reservoir. This 
allows us to study not only the equilibrium states of the quantum chaotic 
environment, but also the relaxation processes towards those states. We believe 
that the non-equilibrium dynamics, and the freedom to choose very strong or 
very weak couplings (in those cases, the canonical ensemble picture is not 
expected to work), open up new and interesting lines of research, where quantum 
chaos may lead to new effects.

One of the simplest examples of an open quantum system with well defined 
canonical thermodynamic properties is the harmonic oscillator coupled to an 
environment which by itself consists of a continuous collection of 
oscillators~\cite{CalLeg83,BrePet02}, a model which is also known under the 
name of ``quantum Brownian motion''. It can be described by the 
Caldeira-Leggett master equation~\cite{CalLeg83,Ram09}. Even though this 
equation is neither exact nor of Lindblad form, it is an excellent 
approximation as long as the temperature is not too low. In this model, the 
central harmonic oscillator evolves asymptotically into the canonical mixture 
of eigenstates with the corresponding Boltzmann weight factors. We then 
introduce quantum chaotic dynamics into the system (the central harmonic 
oscillator) by applying periodic kicks to the oscillator. Without heat bath, 
this system, the quantum kicked harmonic oscillator (KHO)
has been studied in considerable detail; classically in 
Refs.~\cite{Chernikov88} (for an introduction 
see~\cite{Reichl1992,Zaslavsky02} and references therein), and quantum 
mechanically~\cite{GCZ97,CaMaDa04,BilGar09,Kells04}.
The combination of the very simple 
master equation for the harmonic oscillator and the periodic kicks which do not 
interfere with the dissipative dynamics between the kicks has the advantage 
that numerical simulations can be done very efficiently without further 
approximations, and that it can even be realized experimentally~\cite{Lem12}.
The quantum KHO with dissipation has been studied previously in 
Ref.~\cite{CaMaDa04}. However, in that work the authors considered the two 
limiting cases of zero and infinite temperature. Also they concentrated on the
initial stage of the evolution, investigating the ``breaking time'' where the 
quantum evolution starts to deviate notably from the classical one. 

The advantage of introducing quantum chaos with the help of a time-dependent 
potential and not via an additional degree of freedom lies in the reduced 
numerical requirements. The disadvantage lies in the fact that energy is no
longer conserved. Thus we cannot apply standard thermodynamics concepts such as
the canonical ensemble when kicking is present. Though note that the 
thermodynamics of time-periodic systems has been treated 
in~\cite{KetWus10,LanHol14}.

For the simulations we use the Fourier transform of the Wigner function of the 
system which has been called the ``chord function'' by de Almeida and 
coworkers~\cite{Ozo98,Ozo02}. We solve analytically for the chord function 
of the harmonic oscillator in contact with the heat bath and then apply a kick 
to the oscillator. By using interpolation techniques we are able to repeat this 
joint mapping of dissipative dynamics and unitary kicks, which hence yields the
full evolution of the system. We focus on the effects of the coupling to the 
thermal bath in different parameter regimes, such as on and off quantum 
resonances~\cite{BilGar09}, as well as parameter regimes, where the classical 
counterpart changes from integrable to chaotic~\cite{Prado2016}.

The paper is organized as follows: In Sec.~\ref{II} we describe the model and
the method applied to obtain our numerical simulations. Then, in Sec.~\ref{III}
we present our simulations in two parts, the first concentrating on the 
equilibrium properties at relatively strong coupling to the heat bath, the 
second showing the re-appearance of the dynamical properties of the closed 
system, when the coupling to the heat bath is reduced. Finally, in Sec.~\ref{C} 
we present our conclusions.

\section{\label{II} The model}

In this section, we introduce the quantum master equation, which describes the
system of interest (Secs.~\ref{MM} and~\ref{ML}) and then describe our method
to perform the numerical simulations (Secs.~\ref{MD} and~\ref{MN}).

\subsection{\label{MM} Quantum master equation}

Choosing a linear and separable coupling between system and environment and 
restricting oneself to high temperatures, it is possible to derive the 
following quantum master equation
\begin{align}
&\rmi\hbar\, \frac{\rmd\varrho}{\rmd t} =
  [\Hosc,\varrho] + \gamma\; [\hat X,\{ \hat P,\varrho\}]
    -\rmi\, \frac{2m\gamma k_{\rm B} T}{\hbar}\; [\h{X},[\h{X},\varrho]],
\label{CLme}\\
&\text{with}\quad 
\Hosc = \frac{\h{P}^2}{2m} + \frac{m\omo^2}{2}\; \h{X}^2 \; .
\end{align}
This equation has been originally derived by Caldeira and Leggett in 
Ref.~\cite{CalLeg83}, under the assumption that the environment consisted of a 
continuous collection of harmonic oscillators. Here, the central harmonic 
oscillator has mass $m$ and angular frequency $\omo$. The damping constant 
$\gamma$ characterizes its relaxation rate which is related to the Ohmic 
spectral density of the collection of harmonic oscillators in the environment.
Finally, $T$ is the equilibrium temperature of these oscillators, and 
$k_{\rm B}$ is the Boltzmann constant.

In order to add the periodic kicks to the system, we simply replace $\Hosc$ by
\begin{align}
\Hkosc = \Hosc + K\; \cos(\mu \h{X}) \sum_{n\in\mathbb{Z}} 
   \delta(t - n\, T_{\rm k}) \; ,
\end{align}
where the kick wave number is $\mu=2\pi/\lambda_\mt{k}$ with $\lambda_\mt{k}$ 
being the kick wave length. Kick strength and the time period between two kicks 
are denoted by $K$ and $T_{\rm k}$, respectively.

When the kick strength dominates over the coupling to the heat bath, the 
system essentially behaves as if $\gamma = 0$, and we recover the ordinary
KHO where the evolution is unitary~\cite{BilGar09,Kells04}. 
This model has a wide range of dynamical features. To study the different 
regimes, it is simplest to start with the relation between the fundamental 
period of the harmonic oscillator $2\pi/\omo$ and the kick period $T_{\rm k}$. 
Their ratio,
\begin{align}
q= \frac{2\pi}{\omo\, T_{\rm k}} \; ,
\label{MM:q}\end{align}
may be rational or irrational, where the former generally leads to the 
formation of a ``stochastic web'' extending over the whole unbounded phase 
space. In the special cases $q=1,2,3,4,6$ the stochastic web has a crystal 
symmetry, otherwise it forms a quasi-crystal structure. Apart from $q$, 
the system has two additional independent parameters, the kick wave number 
$\mu$ and the kick strength $K$. These two parameters define the overall scale
for the dynamics in phase space, and the degree of chaos. This will be worked
out in more detail in Sec.~\ref{ML}, below. The degree of chaos is understood
to mean the relative size of the areas occupied by the stochastic web vs. the 
islands of integrable motion. While the overall scale does not make a 
difference for the classical dynamics, this is not so in the quantum case. 
There, the size of the primitive crystal cell, may be compared to $\hbar$, 
such that a different relation may lead to different dynamics. This fact is 
the origin for the so called ``quantum resonances'', which lead to quadratic
energy absorption for certain values of the size of the primitive cells, while
otherwise the energy absorption is typically only linear in the number of 
kicks applied to the system. Finally, as first discovered in the kicked rotor,
one may also observe dynamical localization~\cite{Frasca97}.

In the opposite case, when the coupling dominates over the periodic kicks, we
may assume the kick strength to be equal to zero, $\rmK=0$. Then the system 
becomes the quantum harmonic oscillator coupled to a heat bath in the high 
temperature regime~\cite{Ram09,BrePet02}. In this case, the system tends 
towards a thermal equilibrium state, following closely the dynamics of the 
classical damped harmonic oscillator (see App.~\ref{Ap2}).

\subsection{\label{ML} Dimensionless model}

Choosing suitable units for position ($\h{X} = \sqrt{\hbar/(m \omo)}\, \h{x}$) 
and momentum ($\h{P} = \sqrt{\hbar m \omo}\, \h{p}$), the master 
equation~(\ref{CLme}) is rewritten as:
\begin{align}
\rmi\, \frac{\rmd\varrho}{\rmd\tau} = [H_\kappa,\varrho] + \frac{\beta}{2}\;
   [\h{x}, \{\h{p},\varrho\} ] - \rmi\beta\, D\, [\h{x}, [\h{x},\varrho]] \; ,
\label{ML:ME}\end{align}
where energy is measured in units of $\hbar\omo$ such that
$H_\kappa = \hbar\omo\, \Hkosc$ and in terms of the new position and momentum
operators
\begin{align}
H_\kappa = \frac{\hat p^2 + \hat x^2}{2} 
  + \frac{\kappa\, \cos(\sqrt{2}\, \eta\, \hat x)}{\sqrt{2}\, \eta^2}\;
    \sum_{n\in\mathbb{Z}} \delta\Big (\tau - 2\pi\, \frac{n}{q}\Big ) \; ,
\label{ML:Hkappa}\end{align}
where $q$ is the number of kicks per oscillator period, as given in 
Eq.~(\ref{MM:q}). Here, we also introduced the dimensionless time $\tau=\omo t$ 
to describe the evolution of the system. The relative thermal energy becomes 
the dimensionless diffusion constant $D = {k_{\rm B}\, T / (\hbar\, \omo)}$ 
from the quantum Brownian motion model, $\beta= 2\gamma/\omo$ the dimensionless 
energy decay rate. Finally, the so called Lamb-Dicke parameter
$\eta=\mu\sqrt{\hbar/2 m \omo}$ fixes the overall scale of the classical or 
quantum dynamics in phase space, while the effective kick strength 
$\kappa= \mt{K} \mu^2/\sqrt{2} m \omo $ determines the degree of chaoticity in 
the classical system.

\subsection{\label{MD} Dynamics}

In general, the dynamics of the system is given in terms of two alternating
autonomous quantum maps, $\Lambda_\beta$ and $\Lambda'$. The first map
describes the dissipative dynamics under the Caldeira-Leggett master equation,
\begin{align}
\rmi\, \frac{\rmd\varrho}{\rmd\tau} = \frac{1}{2}\; 
   [ (\hat p^2 + \hat q^2) ,\varrho] + \frac{\beta}{2}\;
   [\h{x}, \{\h{p},\varrho\} ] - \rmi\beta\, D\, [\h{x}, [\h{x},\varrho]] \; ,
\label{MD:CalLegEq}\end{align}
for the time $2\pi/q$ between two consecutive kicks. For $\beta > 0$, this
map turns pure state into mixed states which makes it necessary to describe the
whole dynamics in the space of mixed states. The second map is a unitary
transformation, which amounts to an instantaneous change in the momentum wave
function of the system. Thus, for $\varrho(\tau)$ is a solution of the 
Caldeira-Leggett master equation,
\begin{align}
\Lambda_\beta \; :\; \varrho(0) \to \varrho(2\pi/q) \; , \qquad
\Lambda' \; :\; \varrho \to 
   \varrho'= U_\kappa\; \varrho\; U_\kappa^\dagger \; ,
\label{MD:LambdaPrime}\end{align}
where
\begin{align}
U_\kappa =\exp\Big [\, \frac{-\rmi\kappa}{\sqrt{2} \eta^2}\, 
   \cos(\sqrt{2}\eta \h{ x})\, \Big ]\; .
\label{MD:Ukappa}\end{align}
For definiteness, let us agree to start always with the evolution 
$\Lambda_\beta$ under the Caldeira-Leggett master equation. Then, we obtain for
the solution of Eq.~(\ref{ML:ME}) with the initial state $\varrho_0$:
\begin{align}
\varrho_n^+ = \varrho(\tau_n^+) = 
   \big ( \Lambda' \circ \Lambda_\beta \big )^n\; \varrho_0 \; ,
\end{align}
where the product symbol $\circ$ means the composition of the two maps (the 
left one to be applied after the right one), while the $n$'th power means the
repeated composition of $n$ times the same map. This yields the state of the
system right after the $n$'th kick, i.e. at an infinitesimal time lapse 
after the time $\tau_n = 2\pi\, n/q$.

The Caldeira-Leggett master equation can be easily solved in terms of the 
Fourier transform of the Wigner function~\cite{Roy99,BrePet02} 
\begin{align}
\rmw(k,s; \tau)&= \int_{-\infty}^{\infty}\int_{-\infty}^{\infty}\rmd p\, \rmd z
  \; \rme^{\rmi z k+\rmi s p}\, \mt{W}(z,p; \tau) \notag\\
&= \int_{-\infty}^{\infty}\rmd z\, \rme^{\rmi z k}\,
      \la z+s/2 |\,  \varrho(\tau)\,  | z-s/2\ra \; .
\label{Wtransf}\end{align}
This function $\rmw(k,s;\tau)$ also goes under the name of ``chord function'', 
as introduced by de Almeida~\cite{Ozo98,Ozo02}. In the second integral appears 
the mixed state $\varrho(\tau)$ in the position representation 
$\la x|\, \varrho(\tau)\, |x'\ra$ in Dirac notation. The advantage of the chord 
function representation is due to the fact that the Caldeira-Leggett equation 
becomes a first order partial differential equation in the three variables 
$k,s$ and $\tau$,
\begin{align}
\partial_{\tau}\rmw + (\beta s-k)\, \partial_s\rmw
  + s\, \partial_k\rmw = -D\beta\, s^2\, \rmw \; ,
\label{psmasteqap}\end{align}
which can be solved by the method of characteristics. The calculation, worked
out in App.~\ref{Ap1}, yields the following result:
\begin{align}
\rmw(\vec{r},\tau + \sigma) = 
\rmw\left(\, \bM(-\sigma)\; \vec{r},\, \tau  \,\right)
\rme^{-D\beta\left[\, \vec{r}^{T} \, \bA(\sigma)\, \vec{r}\, \right]} \; ,
\label{solme}\end{align}
where a state at time $\tau$ with chord representation 
$\rmw(\vec{r},\tau)= \rmw(k,s;\tau)$ is mapped onto its image at time 
$\tau+\sigma$, with chord representation $\rmw(\vec{r},\tau + \sigma)$. The
matrices $\bM(-\sigma)= \bM(\sigma)^{-1}$ and $\bA(\sigma)$ are given in 
App.~\ref{Ap1}. There, one may convince oneself that $\bM(\sigma)$ simply
describes the classical evolution of the damped harmonic oscillator with 
damping rate $\beta/2$ in phase space. We may thus write for the action of 
$\Lambda_\beta$ in this representation:
\begin{equation}
\Lambda_\beta \; :\; \rmw(\vec{r},0) \to \rmw(\vec{r},2\pi/q) \; ,
\end{equation}
with $\sigma$ in Eq.~(\ref{solme}) replaced by $2\pi/q$.

In order to complete the evolution of the system in the chord representation,
we also need to describe the unitary map $\Lambda'$, in this representation. 
A straight forward calculation, which first switches from the chord function to 
the position representation, then applies the kick according to 
Eqs.~(\ref{MD:LambdaPrime},\ref{MD:Ukappa}), and finally switches back to the 
chord function representation, yields:
\begin{equation}
\Lambda'\; :\; \rmw(\vec{r},\tau) \to \rmw'(\vec{r},\tau) 
 = \sum_{l=-\infty}^{\infty} \mathcal{A}_l(s) \, \rmw(\vec{r}_l,\tauk) \; ,
\label{kick}\end{equation}
where $\vec{r}_l=(k - \sqrt{2}\eta\, l, s)$ and
\begin{equation}
\mathcal{A}_l(s)= (-1)^{l}\; \mt{J}_l\!\Big [ 
   \frac{\sqrt{2}\, \kappa}{\eta^2}\,
   \sin\!\big ( \eta\, s/\sqrt{2}\, \big )\,  \Big ] \; .
\end{equation}
In this expression, $\mt{J}_{l}(z)$ is the Bessel function of the first 
kind~\cite{AbrSte70}. Thus, the effect of the kick consists in generating a
superposition of an infinite copies of the original chord function, each of
which with a specific amplitude and displacement along the variable $k$.

\subsection{\label{MN} Numerical implementation}

The numerical implementation of the evolution of the system relies on the 
ability to accurately represent the true chord function and of the evolving
mixed quantum state, and to accurately implement the two quantum maps 
$\Lambda_\beta$ and $\Lambda'$. Our numerical approach consists of storing
the chord function as a two-dimensional array of function values on a equally
spaced grid in $(k,s)$ space. Then, the application of $\Lambda_\beta$ 
requires to update the function value on each grid point according to the
equation~(\ref{solme}). It is easily seen, that this step requires the 
knowledge of the function values of the original state at points $(k,s)$ in
between the grid points. We estimate these function values with the help of 
the bilinear interpolation method~\cite{NumRec92}. By contrast, the application
of $\Lambda'$ via Eq.~(\ref{kick}) is easier. Since the $s$ variable does not
change, we only need to perform a one-dimensional interpolation on the 
displaced $k$ variable. For simplicity (consistency) we also choose a linear
interpolation scheme in this case, also.

With the chord function at hand, we use separable routines to calculate
probability densities in position and momentum space, as well as expectation
values of the first and second moments of position and momentum. Finally, we
use the two dimensional fast Fourier transform~\cite{NumRec92} to obtain the 
Wigner function representation. The simulations presented in this work are 
performed on a current workstation, using grids with up to 8000 $\times$ 8000 
grid points.

\section{\label{III} Simulations}

In this section we study the effect of the coupling to a heat bath on
the quantum KHO~\cite{BilGar09}. This system has been
thoroughly studied as a closed system, as it as an extremely rich range of 
interesting dynamical features (see discussion elsewhere). However, the open 
system with coupling to a heat bath much less is known; here the only 
treatment, we are aware of is Ref.~\cite{CaMaDa04}.

The present work focuses on the equilibrium properties of the system and the 
validity of thermodynamic concepts in this regime. In Sec.~\ref{SE}, we thus 
start with the case of relatively strong coupling to the heat bath, where we 
expect that specific dynamical properties of the KHO play only a minor 
role, and where the equilibrium states are close 
to the thermodynamic equilibrium states of the harmonic oscillator without 
kicks. In the second part of this section, Sec.~\ref{SD}, we will then reduce 
the coupling to the heat bath, and observe the re-appearance of different 
dynamical effects related to quantum chaos and quantum resonances.

\subsection{\label{SE} Thermodynamical properties at strong coupling}

In the quantum KHO any initial state, normally tends to spread over the whole
phase space reaching an ever higher expectation value $\la\Hosc\ra$. By 
contrast, when the system is coupled to a heat bath, one rather expects that 
eventually, the system's energy saturates at a finite value. Accordingly, the 
evolution of the quantum states becomes periodic at large times, such that with
each kick, the Wigner function expands in phase space (accompanied with an 
increase in energy), and relaxes again towards the harmonic oscillator 
equilibrium state (see Eq.~(\ref{wth}) in App.~\ref{Ap2}) afterwards, until it 
comes to the next kick. While this behavior is expected to happen for any 
finite coupling, it is difficult to observe when the coupling is small. This is 
because then, the regime of cyclic behavior is reached only at large time and 
the system average energy is large, also. In this section, we choose a rather 
large coupling $\beta= 0.1$, such that the cyclic regime is reached rather 
quickly and easily observable.

\begin{figure}
\includegraphics[width=0.5\textwidth]{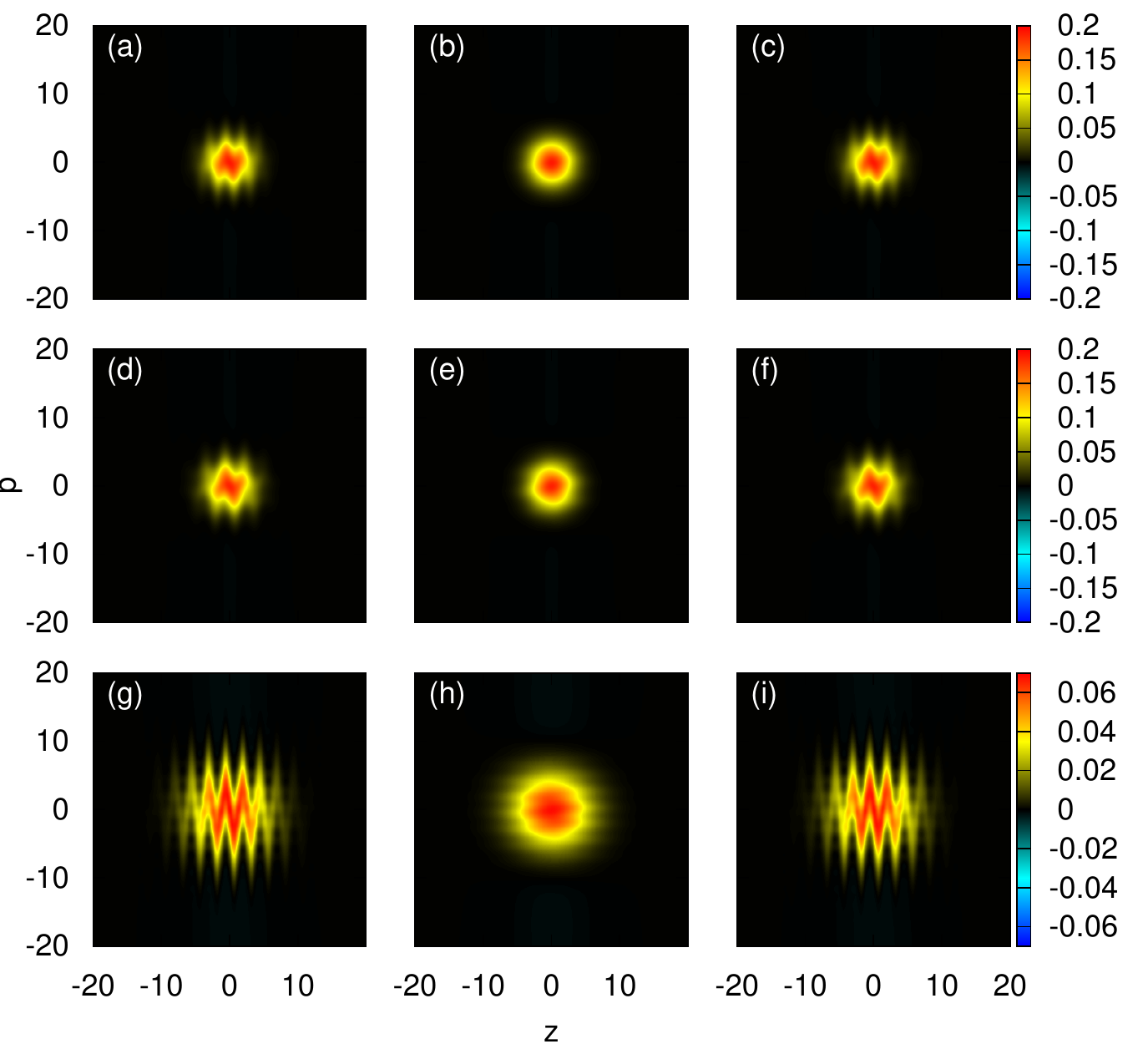}
\caption{Wigner function for the quantum KHO for $q=4$, coupled to a heat bath 
(with $\beta=0.1$ and $D=5$), with an initial coherent state positioned at 
$z=0, p=0$. Different rows correspond to different choices of $\kappa$ and 
$\eta$. Resonant case (first row, $\kappa= -0.8, \eta^2= \pi$) after the 35'th
kick (a), before the 36'th kick (b) and after the 36'th kick (c). Non-resonant 
case (second row, $\kappa= -0.8, \eta^2= 0.7\, \pi$) at the same 
instances in time. Chaotic case (third row, $\kappa= -4.5, \eta^2= 1$).}
\label{fig1}\end{figure}

\begin{figure}
\includegraphics[width=0.5\textwidth]{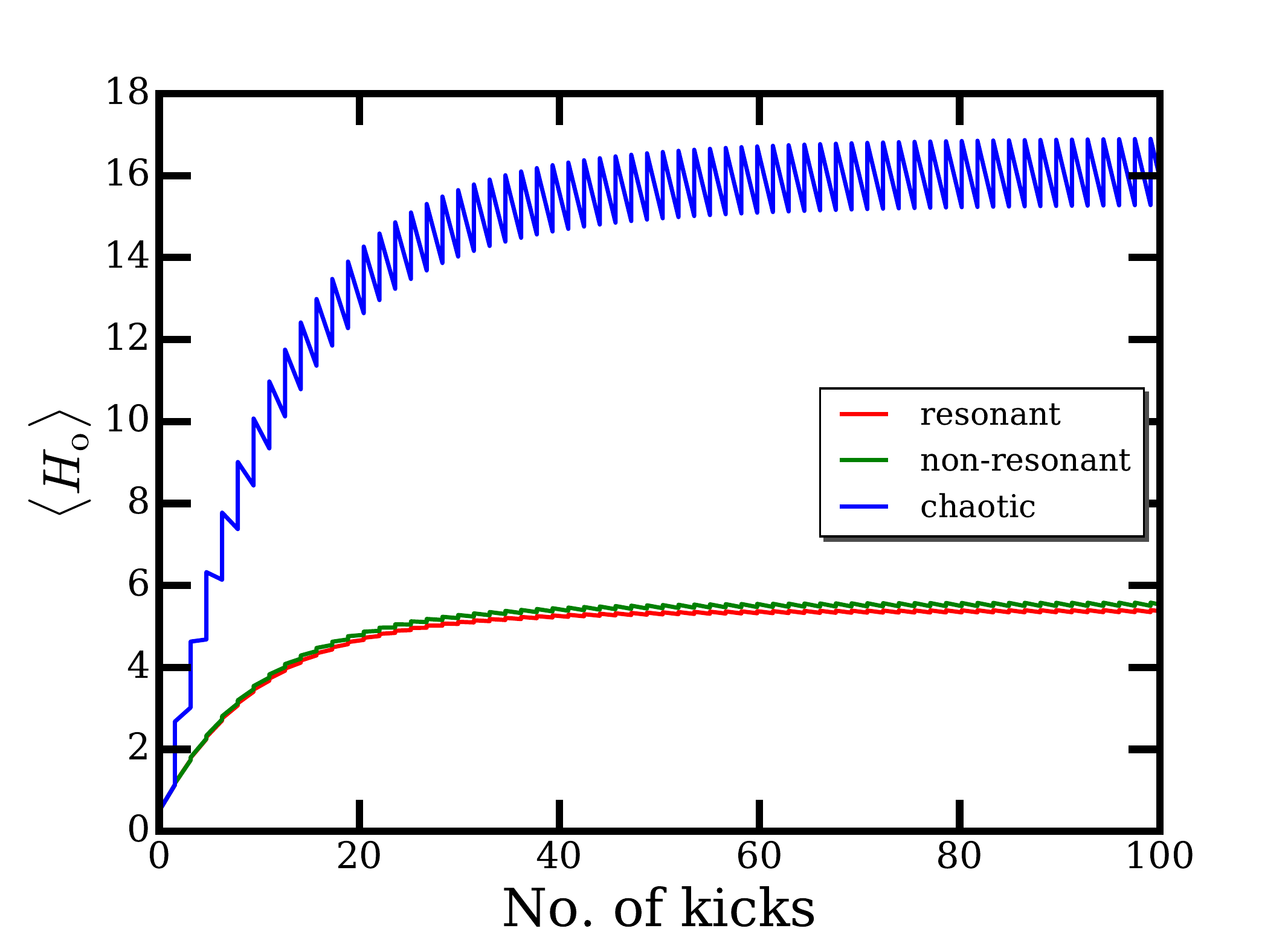}
\caption{Energy of the harmonic oscillator $\la\Hosc\ra$ vs. the number of 
kicks for the same three cases, shown in Fig.~\ref{fig1}. The lowest curve
(red line) corresponds to the resonant case with $\kappa=-0.8$ and 
$\eta^2=\pi$; the curve slightly above (green line) to the non-resonant case
with $\kappa=-0.8$ and $\eta^2= 0.7\, \pi$. The top curve (blue 
line) corresponds to the chaotic case with $\kappa=-4.5$ and $\eta^2=1$.}
\label{fig2}\end{figure}

In Fig.~\ref{fig1} we present Wigner functions of an evolving quantum states,
starting out at $\tau= 0$ from the coherent state in the center of the phase
space. We show the Wigner functions just after the 35'th kick (first row), 
right before the 36'th kick (second row) and right after the 36'th kick (third
row). At these times, the system is already very close to its limit cycle 
behavior in all the cases, considered. This can be seen from Fig.~\ref{fig2},
below. The main purpose of the figure is it to show that the limit
cycle behavior is practically independent on the dynamical regime of the 
isolated KHO. The only quantity, which really matters is the kick amplitude 
$\kappa/(\sqrt{2}\, \eta^2)$ from Eq.~(\ref{ML:Hkappa}) which determines the
amount of energy transfered to the system at each kick.

The first row shows the resonant case, with $\kappa= -0.8$ and $\eta^2= \pi$
such that the kick amplitude is $\kappa/(\sqrt{2}\, \eta^2) \approx 0.18$. The
second row shows the non-resonant case, with $\kappa= -0.8$ and 
$\eta^2= 0.7\, \pi$, where 
$\kappa/(\sqrt{2}\, \eta^2) \approx 0.25$. The third row 
shows the chaotic case, with $\kappa= -4.5$ and $\eta^2= 1$, where 
$\kappa/(\sqrt{2}\, \eta^2) \approx 3.18$. Remember, while $\kappa$ determines
the degree of classical chaos in the system, $\eta$ is a scale factor which
determines the size of classical structures on the phase space where quantum 
states occupy on average an area of size one. Thus, the semiclassical limit 
(usually denoted as $\hbar\to 0$) corresponds here to the limit 
$\eta^2\to \infty$. 

Comparing the Wigner functions shown in the first and the second row, we can
hardly note any differences. This confirms the dominant effect of the coupling
to the heat bath. Without coupling, i.e. for the closed KHO, we would expect a 
much more extended Wigner function in the resonant case (here, the energy of
the system increases quadratically in time) than in the non-resonant case 
(here, it increases only linearly). The Wigner functions in the third row
correspond to the chaotic chase. There, the Wigner functions are much more 
extended. However, comparing the kick amplitudes calculated in the previous
paragraph ($3.18$ for the chaotic case, vs. $0.18$ and $0.25$ for the resonant
and non-resonant cases) we find that this larger extension is mainly due to
the kick amplitude. The zig-zag pattern, most clearly recognizable for the 
states right after a kick can be directly related to the kick potential, as its 
periodicity in the z-direction agrees with that of the kick potential.

In Fig.~\ref{fig2}, we show the evolution of the system energy $\la\Hosc\ra$
as a function of the number of kicks, for the same three cases depicted in 
Fig.~\ref{fig1}. Since $q=4$, the number of kicks which amount to one oscillator
period (duration $2\pi$ in our dimensionless units) is equal to four. As we
chose the same initial state (a coherent state at the origin), the energy of the
system starts out at $\la\Hosc\ra$. As can be seen from the figure, the 
evolution starts out at $\tau = 0$ with a solution of the Caldeira-Leggett 
master equation. Since the energy of the system is smaller than the thermal 
energy of the heat bath, the system absorbs energy from the heat bath. Later on,
the system energy will always be larger and the system will release energy to
the heat bath. We observe that the two curves for the resonant and the 
non-resonant case are very close together, and they reach approximately the
same final average energy, close to $D=5$. This means that the effect of the
kicks is weak as compared to the heat bath. As a consequence, the limit cycle
states are close to the thermal equilibrium state with $\la\Hosc\ra = D = 5$.

In the chaotic case (top most blue line), one can clearly observe the effect of
each individual kick, and the subsequent relaxation. Here, the effect of the
kicks is strong as compared to the heat bath. As a consequence, the average
energy of the system is much larger than it would be due to the coupling to 
the heat bath alone. One may suspect that the limit cycle states vary around a 
thermal state with average energy close to $\la\Hosc\ra = D = 16$.

\begin{figure}
\includegraphics[width=0.5\textwidth]{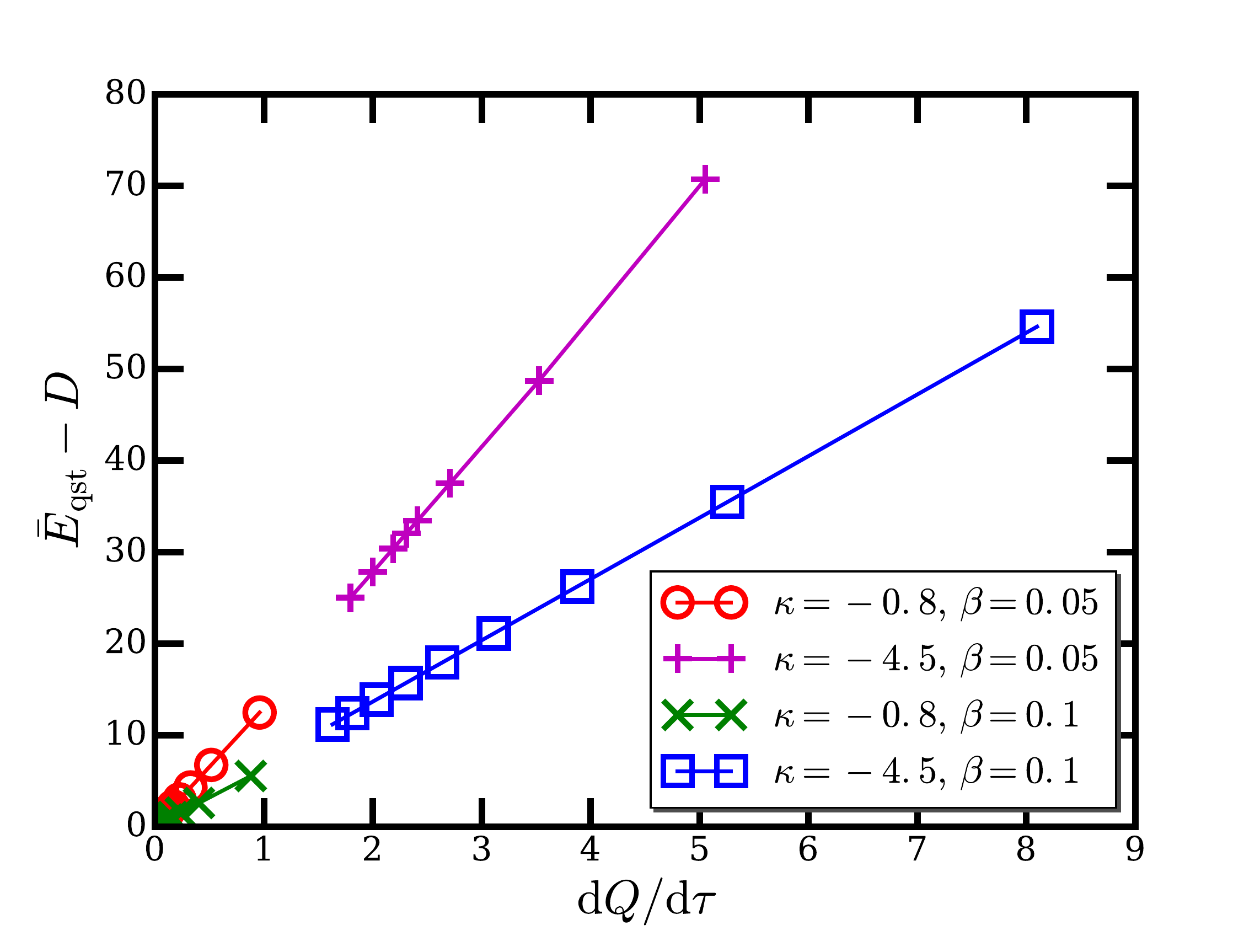}
\caption{Difference between average system energy $\Eqst$ from 
Eq.~(\ref{SE:Eqst}) vs. the heat current $\rmd Q/\rmd\tau$ from 
Eq.~(\ref{SE:DeltaQ}), for different KHO parameters $\kappa$ and $\eta^2$, as 
well as coupling strengths $\beta$; see legend. For each of the four cases, 
$\eta^2/\pi$ varies from $0.1$ to $1$ in integer steps. Larger values of 
$\eta^2$ mean smaller kick amplitudes, and hence the corresponding points are 
further away from the origin.}
\label{fig3}\end{figure}

Let us now turn our attention to the energy balance in the 
quasi-stationary regime. Specifically, we consider the energy of the harmonic 
oscillator as a function of time, $E(\tau)= \la\Hosc\ra$ at the time $\tau$. 
As far as time is concerned, we make use of the notation introduced in
Sec.~\ref{MD} we denoted the time right after the $n$'th kick by 
$\tau_n^+ = 2\pi\, n/q + \delta$ with an infinitesimal increment $\delta$ (be
reminded that time is measured in units such that one oscillator period is
equal to $2\pi$). Similarly, we now denote $\tau_n^-$ as the time right before 
the $n$'th kick. The quasi-stationary regime may thus be characterized by the
condition that the whole energy gained from any one kick, 
$E(\tau_n^+) - E(\tau_n^-)$ is subsequently transferred to the heat bath during 
the time interval $(\tau_n^+,\tau_{n+1}^-)$. For this regime, we hence 
introduce the average quasi-equilibrium energy of the oscillator as 
\begin{align}
\Eqst = \lim_{n\to\infty} \frac{E(\tau_n^+) + E(\tau_n^-)}{2} \; .
\label{SE:Eqst}\end{align}
On a coarse grained time scale of the order of many cycles, we may interpret
the dynamics as a thermodynamic process where work (energy transfer due to the
kicks) is continuously converted into heat, released to the reservoir. On the
above mentioned coarse grained time scale, the heat flux from the oscillator to
the heat bath is constant, and given by
\begin{align}
\frac{\rmd Q}{\rmd\tau} = \frac{q}{2\pi}\; 
   \lim_{n\to\infty} \big [ E(\tau_n^+) - E(\tau_{n+1}^-) \big ]\; .
\label{SE:DeltaQ}\end{align}
In Fig.~\ref{fig3}, we study the amount of heat transfered to the reservoir as 
a function of the energy difference $\Eqst - D$, which may be interpreted as a
temperature difference between the heat bath and the oscillator. This assumes
that it is possible to assign a temperature to the oscillator in this quasi-
equilibrium situation. One then expected that the Fourier law (of thermal
conduction) would hold, which predicts a linear relation between both 
quantities. Indeed, Fig~\ref{fig3} seems to confirm the validity of the Fourier 
law.

\subsection{\label{SD} Weak coupling and reappearance of quantum 
  chaotic/resonance properties}

In the previous Sec.~\ref{SE}, we studied the case of strong coupling to the
heat bath, where different dynamically properties of the quantum KHO almost do
not play any role. It is then natural to ask, at which scales and how the
different dynamical features re-appear, when the coupling to the heat bath is
reduced. This is the purpose of the present section.

For the KHO, we have essentially three different parameters which can be 
changed, (i) the kick period $\tau_{n+1} - \tau_n = 2\pi/q$, which may or may 
not be commensurable with the period $2\pi$ of the harmonic oscillator; (ii)
the scale-invariant kick strength $\kappa$, which determines the degree of 
chaos in the corresponding classical system; and (iii) the scale (size) of the 
system in phase space, which is determined by $\eta$. As mentioned earlier,
taking $\eta\to\infty$ implements the semiclassical limit. However, $\eta$ is
also the parameter which controls the quantum resonances, as it happens that
for specific integer values of $q$, the system energy can increase 
quadratically in time, whereas otherwise it only increases linearly. 
For small coupling $\beta$, the study of the system is limited to a relatively 
narrow range of parameters or to short times. This is because, our numerical 
scheme starts to fail if the Wigner functions become too extended in phase 
space. In those cases, it is increasingly difficult to reach the 
quasi-stationary regime.

\begin{figure}
\includegraphics[width=0.5\textwidth]{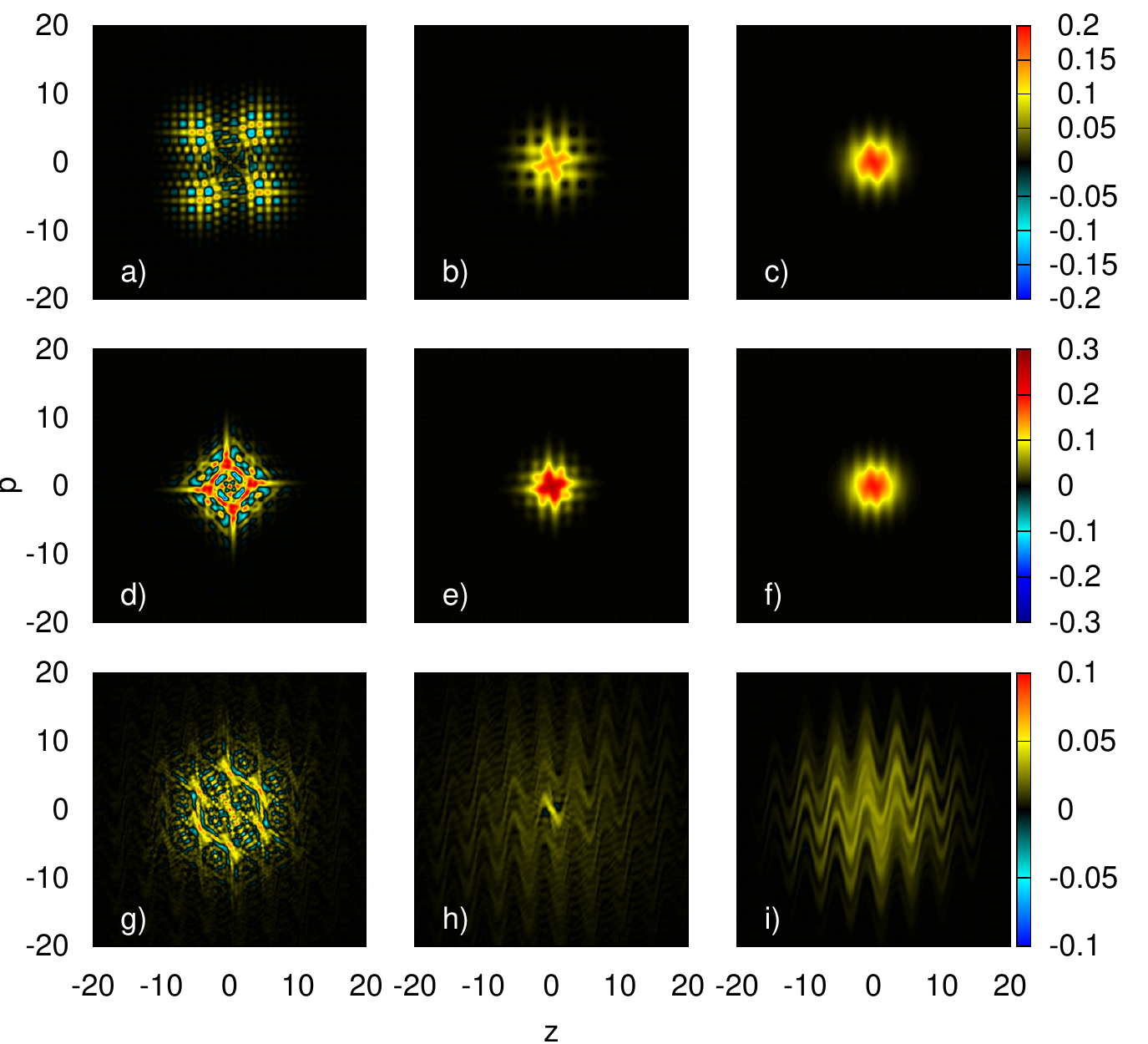}
\caption{Wigner function for the KHO coupled to heat bath, right after the 
  36'th kick, for different coupling strengths: $\beta=0.001$ (first column), 
  $\beta=0.01$ (second column), and $\beta=0.1$ (third column). Again we 
  consider three cases, the resonant case (first row) with $\kappa=-0.8$ and 
  $\eta^2=\pi$; the non-resonant case (second row) with $\kappa=-0.8$ and 
  $\eta^2=(1+\sqrt{5})\, \pi/2$ (resonant cases is considered in
  the Figs.~\ref{fig1} and~\ref{fig2}). The chaotic case (third row) with
  $\kappa=-4.5$, $\eta^2=1$, however has a different inter-kick period, with
  $q=6$.}
\label{fig4}\end{figure}

In Fig.~\ref{fig4} we show Wigner functions after the 36'th kick. Similar
to Fig.~\ref{fig1}, we consider the resonant, non-resonant and chaotic case,
each case in a different column. However, we now vary the coupling strength
from $\beta= 0.001$ (first column) to $\beta= 0.01$ (second column), and 
$\beta= 0.1$ (third column). The cases shown in the third column agree with
those shown in Fig.~\ref{fig1}, except for the chaotic case, where we chose
$q=6$ instead of $q=4$. In this case, where $\beta$ is sufficiently large, the
quasi-stationary regime is almost reached, and we can observe some minor 
differences between the $q=6$ and $q=4$ case. For small values of $\beta$
(first column), we find regions where the Winger function becomes negative.
This is a clear signature for the state to be non-classical. For $\beta= 0.001$
and $0.01$, we also observe that the extension of the Wigner function is larger 
for the resonant case than for the non-resonant case, as expected according to
the quadratic over linear energy absorption. As shown in Fig.~\ref{fig2}, this
is no longer the case for $\beta=0.1$, where the quantum resonance condition
seems to become meaningless. In the chaotic case, the extension of the Wigner
function is by far largest. However, as explained in Fig.~\ref{fig1}, this is
a simple consequence of the large kick amplitude. In the chaotic case, one
would expect a linear increase in energy, just as in the non-resonant case
~\cite{Kells04}. Note that the hexagonal symmetry 
observable for $\beta= 0.001$, which is due to the choice of $q=6$, disappears
as $\beta$ is increased, until at $\beta= 0.1$ the characteristic zig-zag
shape appears, similar to the $q=4$ case, shown in Fig.~\ref{fig1}.

\begin{figure}
\includegraphics[width=0.5\textwidth]{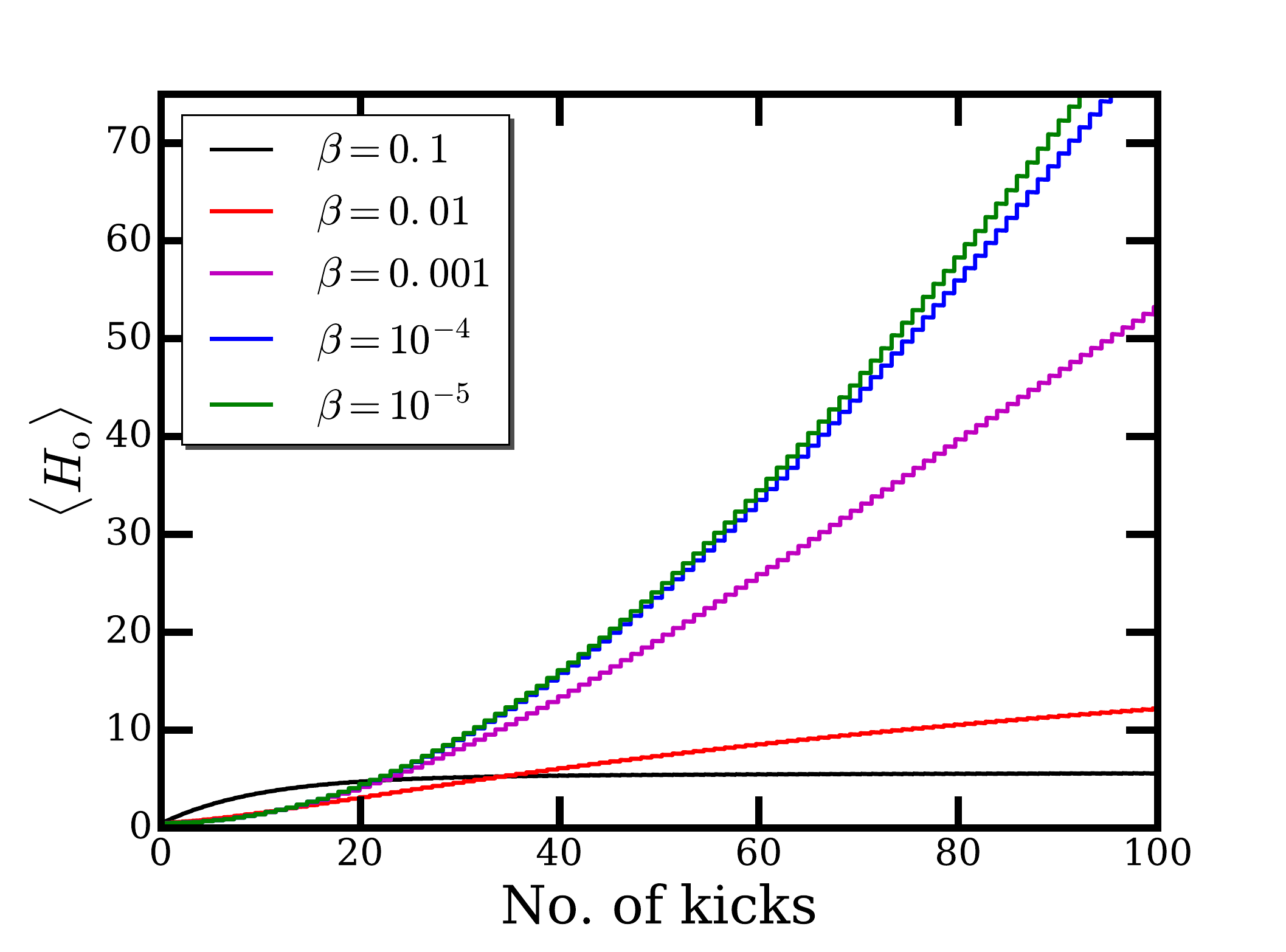}
\caption{Energy evolution in time in the resonant case, 
  $q=4, \kappa=-0.8, \eta^2=\pi$ for different coupling strengths, from 
  $\beta= 10^{-5}$ (green line) to $\beta=0.1$ (black line).}
\label{fig5}\end{figure}

In Fig.~\ref{fig5} the expectation value of the oscillator energy is shown as
a function of the number of kicks applied to the system. The parameters for the
KHO ($q=4, \kappa=-0.8, \eta^2=\pi$) are chosen to be the same as for one of 
the two resonant cases shown in Fig.~3(b) of Ref.~\cite{BilGar09}, where the 
quantum KHO is treated without dissipation. Indeed, our result for 
$\beta= 10^{-5}$ seems to agree very well with the result shown there. We can
clearly see the quadratic increase in energy, which becomes only slightly 
diminished when $\beta$ is increased to $\beta= 10^{-4}$. By contrast, for the
largest coupling, $\beta= 0.1$, the energy only increases in an initial phase,
and then quickly approaches its limit (average) value of $\la\Hosc\ra = D = 5$.
For intermediate couplings, $\beta = 0.001$ and $\beta = 0.01$, the increase of
the energy is no longer quadratic, but the quasi stationary regime only sets in
after many more kicks than could be shown here.

\begin{figure}
\includegraphics[width=0.5\textwidth]{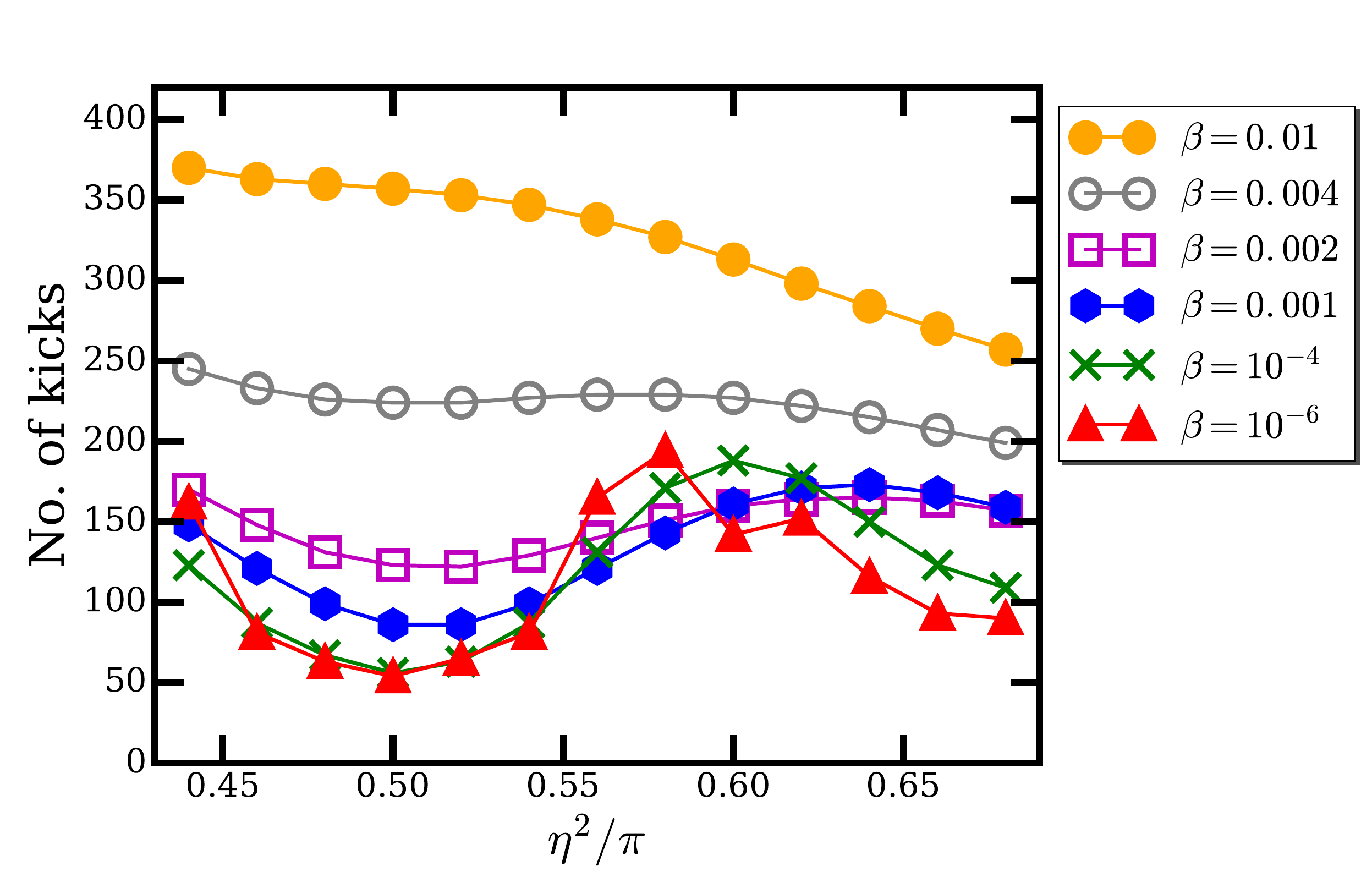}
\caption{Plots of the number of kicks required to reach a mean 
  energy of $50\hbar\omega$ for the resonant case (with $\kappa=-0.8$, $q=4$) and different coupling strength values, 
  $\beta$; the resonance is in $\eta^2/\pi=0.5$.}
\label{fig6}\end{figure}

In Fig.~\ref{fig6} we plot the number of kicks required to reach the energy of 
$\la\Hosc\ra = 50$ (in units of $\hbar\omo$). Again, this is done for the 
resonant case, with $q=4$, $\kappa=-0.8$. This time $\eta$ is varied in a 
small region around the quantum resonance condition $\eta^2 = \pi/2$. This 
corresponds to the second case of a quantum resonance, also considered in 
Ref.~\cite{BilGar09}. For $\eta^2$, we focus on a small section from a similar 
figure from that reference. Now, we plot the number of kicks required to reach 
the above mentioned energy limit, as a function of $\eta^2$ for different 
coupling strengths, from $\beta = 10^{-6}$ to $\beta = 0.01$. For the smallest
coupling strength, we find the expected minimum, and we again confirm a good
agreement with the corresponding result in Fig.~3(a) from Ref.~\cite{BilGar09}.
For increasing $\beta$, the minimum remains at its place, approximately until 
$\beta = 0.002 - 0.004$, and then disappears. At the largest coupling, 
$\beta = 0.01$, the number of kicks required to reach the energy limit becomes
a monotonous function of $\eta^2$. It is noteworthy that increasing the 
coupling from $\beta = 10^{-6}$ to $\beta = 10^{-5}$ already leads to notable
differences, but only sufficiently far away from the quantum resonance. It is
also interesting that for fixed $\eta^2 \approx 0.0575 \pi$, the number of
kicks required does not increase monotonously with $\beta$. This phenomenon of
increasing energy absorption due to increasing the coupling to a heat bath 
might be related to destroying a dynamical localization effect by increasing
decoherence~\cite{Frasca97,Kells04}.

\section{\label{C} Conclusions}

In this paper, we described the quantum kicked harmonic oscillator as an
open quantum system coupled to a finite temperature heat bath. We studied its
equilibrium properties at relatively strong coupling, and found that there, the 
system fulfills fundamental thermodynamic properties such as the Fourier law 
for heat transport. When reducing the coupling to the heat bath, the system's 
equilibrium state (i.e. its Wigner function representation) becomes ever more 
extended in phase space, and the expectation value of the oscillator energy
increases. This makes it ever more difficult to perform simulations for a long 
time (many kicks), and eventually we are no longer able to reach the 
equilibrium state. Thus, for small coupling, we restrict ourselves to study the 
effect of quantum resonances and how it becomes suppressed and eventually 
eliminated due to the increasing coupling to the heat bath. 

The numerical method is limited essentially by the requirement of an accurate
representation of the chord function. At the present stage, we use a simple 
uniform two dimensional grid together with bilinear 
interpolation~\cite{NumRec92}. However, we are confident to be able to improve
that scheme in order to be able to reach higher energies and larger times.

As shown in Ref.~\cite{Lem12} the model is realizable experimentally in an 
quantum optical setup. Alternatively one might think of ions in a harmonic 
trap. The coupling of the ion movement to its internal electronic states opens 
a way to consider the KHO with heat bath as the environment for a central 
quantum system, which may serve as a probe for the thermodynamical properties 
of KHO.

\acknowledgments

We thank A. Eisfeld and F. Leyvraz for enlightening discussions. We acknowledge
the hospitality of the Centro Internacional de Ciencias, UNAM where some of the
discussions took place. We are grateful for the possibility to use the Computer
cluster at the Max-Planck Institute for the Physics of Complex Systems, where
some of the numerical calculations have been performed. Finally, we 
acknowledge financial funding from CONACyT through grant No. CB-2009/129309
at an early stage of the project.

\appendix

\section{\label{Ap1} Solution of the Caldeira-Leggett master equation for 
the quantum harmonic oscillator}

The Caldeira-Leggett master equation for the harmonic oscillator transformed 
into the chord function representation in the dimensionless variables
description reads:
\begin{equation}
\partial_{\tau}\rmw+(\beta s-k)\partial_s\rmw
+s\partial_k\rmw=-D\beta s^2\rmw 
\end{equation}
where $\rmw=\rmw(\vec{r},\tau)$ and $\vec{r}=(k,s)$. This is a first order partial differential 
which can be solved by standards procedures. The equation 
(\ref{psmasteqap}) can be written in its parametric 
form:
equation in the parametric form is 
\begin{equation}\label{ecpar}
 \frac{\rmd s}{\rmd \tau}=\beta s-k,\quad \frac{\rmd k}{\rmd \tau}=s,\quad 
 \frac{\rmd}{\rmd \tau}\rmw(\vec{r},\tau)=-D\beta s^2\rmw(\vec{r},\tau).
\end{equation}
By coupling the first two set of equations, one has
the solution for the case $\beta < 1 $:
\begin{eqnarray}\label{eck1}
k_{\tau}&=&u_1(\tau)a_1+u_2(\tau)a_2\\\label{eck2}
s_{\tau}&=&u_3(\tau)a_1+u_4(\tau)a_2
\end{eqnarray}
where we have called for simplicity 
\begin{eqnarray}
u_1(\tau)=\rme^{\beta \tau \over 2}\sin\omega \tau,&& \quad u_2(\tau)=\rme^{\beta \tau \over 2}\cos\omega \tau\\
u_3(\tau)= \frac{\rmd}{\rmd \tau } u_1(\tau), && \quad u_4(\tau)={\rmd\over \rmd \tau } u_2(\tau),
\end{eqnarray}
with $\omega=\sqrt{1-\beta^2/4}$.
The constants $a_1$ and $a_2$  are known as the characteristics curves and they emerge 
as integration constants of the solution of the first two set of equations in (\ref{ecpar}).
In terms of them, the map of any set of points $k$ and $s$ at the 
time $\tau$ described in terms of the characteristics:
$k=u_1(\tau)a_1+u_2(\tau)a_2$ and $s=u_3(\tau)a_1+u_4(\tau)a_2$
to the points $k'$ and $s'$ at the time $\tau + \sigma$, is  done by
moving along the characteristic curves. This map can be represented 
by the following transformation: 
\begin{equation}
 \vec{r}_{\tau + \sigma } = \bM(\sigma)\; \vec{r}_{\tau}
\end{equation}
where the matrix $\bM(\sigma)$ has the form:
\begin{equation}\label{mmat}
\bM(\sigma)=\rme^{\beta\sigma/2}\left( \begin{array}{cc}
m_1(\sigma) & m_2(\sigma) \\
m_3(\sigma) & m_4(\sigma) 
\end{array}\right)
\end{equation}
and 
\begin{eqnarray}\label{etas}
m_1(\sigma)&=&\cos\omega\sigma-{\beta\over 2 \omega}\sin\omega \sigma,\\
m_2(\sigma)&=&{1\over \omega}\sin\omega \sigma,\\
m_3(\sigma)&=&{-1\over \omega}\sin\omega \sigma,\\
m_4(\sigma)&=&\omega\cos\omega \sigma+{\beta\over 2}\sin\omega \sigma.
\end{eqnarray}
The mapping is reversible as one can check that the relations $\bM(\sigma)^{-1} = \bM(-\sigma)$
holds. The matrix $\bM(\cdot)$ is also composed 
by a rotational and contracting part, so as the maps goes to infinity, the characteristics 
moves any initial set of points into the origin. Integration of the third
equation of (\ref{ecpar}) is of the form
\begin{equation}\label{int3}
\int_{\rmw(\tau)}^{\rmw(\tau+\sigma)}{\rmd\rmw\over \rmw}=-D\beta\int_{\tau}^{\tau+\sigma}\rmd\tau '\, s^2(\tau ') .   
\end{equation}
where $\rmw(\tau)=\rmw(\vec{r},\tau)$ is the initial condition
of the oscillator whose dependence 
on the variables $k$ and $s$ at $\tau$.
By using the relation of the characteristics (\ref{eck1}) and (\ref{eck2}) 
one can write the argument of the integral in the r.h.s of (\ref{int3}) in terms 
of the variables $k$ and $s$ at time $\tau$ 
as: $s(\tau') = \rme^{-{\beta\over 2}
{\left(\tau-\tau'\right)}}\left(m_3(\tau-\tau') k + m_4(\tau-\tau')s\right)$
and perform the integration yielding the solution for the chord function at the 
time $\tau + \sigma$:
\begin{equation}\label{solmeap}\nonumber
\rmw(\vec{r},\tau+\sigma) = 
\rmw\left(\, \bM(-\sigma)\; \vec{r},\, \tau  \,\right)
\rme^{-D\beta\left[\, \vec{r}^{T} \, \bA(\sigma)\, \vec{r}\, \right]}
\end{equation}
where the matrix $\bA(\sigma)$ depends also on the time of propagation from 
the initial condition to the final time and has the form:
\begin{equation}\label{almat}
 \bA(\sigma)=\left(\begin{array}{cc}
A_1(\sigma) & A_2(\sigma) \\
A_2(\sigma) & A_3(\sigma) 
\end{array}\right),
\end{equation}
and
\begin{eqnarray}
A_1(\sigma)&=& \int_{0}^{\sigma} 
\rmd \tau'\, \rme^{\beta \tau'}  m^2_3 (-\tau')\\
A_2(\sigma)&=& \int_{0}^{\sigma}\! \!\rmd \tau'\, 
\rme^{\beta \tau'}\! m _3(-\tau') m _4 (-\tau')\\
A_3(\sigma) &=& \int_{0}^{\sigma} \rmd \tau'\, \rme^{ \beta  \tau'} m^2_4(-\tau').
\end{eqnarray}

\section{\label{Ap2} The stationary state of the quantum harmonic oscillator
in the Caldeira-Leggett model}

As the system propagates in contact with the thermal bath, 
the system relaxes into a thermal stationary state 
correspondent to the Caldeira-Leggett model for which the high temperature 
limit is assumed. The form of the stationary state can be obtained 
by taking the long time limit $\sigma \rightarrow \infty$ in the solution 
of the master equation in the chord function description (\ref{solmeap}).
In this limit, $\lim_{\sigma \rightarrow \infty} \, \bM(-\sigma)=0$. 
If one assumes in a general context an initial condition for the system as coherent state: 
$\psi(X)=({m\omo\over \pi \hbar})^{1/4}\rme^{\rmi P_{\mt o}X - {m\omo\over 2\hbar}(X-X_{\mt o})^2 }$
which in the chord function representation the initial 
condition takes the form:
\begin{equation}\label{icw}
\rmw\left( \vec{r},\tau\right)= \sqrt{m\omo \over 2\pi \hbar }\, \rme^{\rmi {P_{\mt o}\over k_{\mt o}} s+\rmi X_o k -{k^2\over 4} - {s^2\over 4}},
\end{equation}
then in the long time limit, the 
dependence of the system on the state of the initial condition vanishes, yielding only
a constant term:
\begin{equation}
\lim_{\sigma\rightarrow \infty}
\rmw\left(\, \bM(-\sigma)\; \vec{r},\, \tau  \,\right)= \sqrt{m\omo \over 2\pi \hbar }.
\end{equation}
On the other hand the form of the $ \bA(\sigma)$ matrix as $\sigma\rightarrow \infty$ is:
\begin{equation}\label{almatst}
 \lim_{\sigma\rightarrow \infty} 
 \bA(\sigma)={1\over 2 \beta}\left(\begin{array}{cc}
1 & \beta   \\
\beta  &  1 + \beta^2
\end{array}\right),
\end{equation}
which in the limit of applicability of the Born-Markov approximation, $\beta\ll \pi D$,  which is an assumption 
of the derivation of the Caldeira-Leggett master equation, the  
$\bA(\sigma\rightarrow\infty)$ matrix can be approximated to ${1\over 2 \beta} \One$, where $\One$ is the
$2\times2$ unit matrix. The chord function at the stationary regime can be written as:
\begin{equation}\label{solmeapst}\nonumber
\rmw(\vec{r},\sigma \rightarrow\infty) = \sqrt{m\omo \over 2\pi \hbar }
\rme^{-{D\over 2} \left( k^2   +  s^2\right)}
\end{equation}
Transformation of the chord function (\ref{solmeapst}) into the position representation
yields the following  density matrix for the stationary regime:
\begin{equation}
 \varrho_{\mt {ths}}(X,X')=\sqrt{ m \omo^2 \over 2\pi  k_B T }\rme^{-{ m \omo^2 \over 8 k_B T}(X+X')^2- {m k_B T \over 2 \hbar^2}\left(X-X'\right)^2},
\end{equation}
which agrees with the solution for the Caldeira-Legget master equation for the stationary regime, see \eg~\cite{Ram09,BrePet02}
The average energy of the system  at the stationary regime
becomes exactly $ E_{\mt{therm}} = \la H_{\mt{osc}}\ra= k_B T $ as in a thermalization process. The role of 
$\beta$, is related to the rapidity of the relaxation process into the thermal state. The Wigner function of the system at the thermal 
state has the following form:
\begin{equation}\label{wth}
\mt{W}_{\mt {ths}}(z,p)= \sqrt{ \frac{m \omo}{2\pi  \hbar D^2} }\;
\exp\!\Big ( - \frac{z^2+ p^2}{2 D^2}\Big ) \; ,
\end{equation}
which is a Gaussian function centered at the origin with standard deviation $D$.


%
\end{document}